\documentclass[twocolumn,showpacs,preprintnumbers,amsmath,amssymb]{revtex4}

\usepackage{graphicx}
\usepackage{dcolumn}
\usepackage{bm}
\usepackage[latin1]{inputenc}     

\usepackage{float}   

\begin{document}

\title{A Crucial Experiment To Test The Broglie-Bohm Trajectories For
Indistinguishable Particles.}
\author{Michel Gondran}
 \affiliation{EDF, Research and Development, 92140 Clamart.}
 \email{michel.gondran@chello.fr}   
\author{Alexandre Gondran}
 \affiliation{SET, Université de Technologique de Belfort-Montbéliard}
 \email{alexandre.gondran@utbm.fr}   

\date{\today}

\begin{abstract}

The standard quantum theory has not taken into account the size of
quantum particles, the latter being implicitly treated as material
points. The recent interference experiments of
Zeilinger~\cite{Zeilinger_1999} with large molecules like
fullerenes and the thought experiments of Bozic \textit{et
al}~\cite{Bozic_2004} with asymmetrical Young slits make it
possible today to take into account the particle size.

We present here a complete study of this phenomenon where our
simulations show differences between the particles density after
the slits and the modulus square of the wave function. Then we
propose a crucial experiment that allows us to reconsider the
wave-particle duality and to test the existence of the
Broglie-Bohm trajectories for indistinguishable particles.

\end{abstract}

\maketitle

\section{Introduction} \label{}

\bigskip

The standard quantum theory of interference phenomena does not
take the size of the particles into account, the latter being
implicitly treated like material points.  Two significant advances
make this possible now thus allowing us to better apprehend the
wave-particle duality.

The first advance concerns the interference experiments realized
some years ago with large size mesoscopic individual quantum
objects, cf. Schmiedmayer \textit{et al}~\cite{Schmiedmayer_1993}
and Chapman \textit{et al}~\cite{Chapman_1995} with the molecules
of $Na_2$ ($\sim 0.6 nm$ size), Arndt \textit{et
al}~\cite{Zeilinger_1999} with the fullerene molecules $C_{60}$
($\sim$ 1 nm diameter), Nairz \textit{et al}~\cite{Zeilinger_2000}
with the molecules $C_{70}$ and more recently Hackermüller
\textit{et al}~\cite{Zeilinger_2003} with the molecules of
fluorofulleres $C_{60}F_{48}$.

The second advance concerns the thought experiments suggested and
simulated by Bozic \textit{et al}~\cite{Bozic_2002,Bozic_2004}:
those are interference experiments with slits of various sizes,
some large enough to let the molecules get through, others smaller
making that impossible.  It is thus theoretically possible to take
into account the size of the particles~\cite{Bozic_2004} by
studying their differences in behavior according to the respective
sizes of the particles and the slits.

These thought experiments are particularly suggestive concerning
the interpretation of the wave function since they correspond to
cases in which \textit{the particles density measured after the
slits can be different from the calculation of the modulus square
of the wave function}.  It is especially the case when the
particle diameter is larger than the size of all the slits and as
Arndt \textit{et al}~\cite{Zeilinger_1999} underlines it \textit{"
it would be certainly interesting to investigate the interference
of objects the size of which is equal or even bigger than the
diffraction structure".} Indeed, in this case the particles
density after the slits will then be null while the standard
calculation of the modulus square of the wave function will not:
\textit{the postulate of the probabilistic interpretation of the
modulus square of the wave function could well be questioned by
this experiment and must thus be reappraised}.

The initial point of this article is to make a complete study of
this phenomenon by calculating the particles density after the
slits according to the various possible assumptions. The second
point is to propose a crucial thought experiment to test the
existence of the Broglie-Bohm trajectories for indistinguishable
particles. This will be achieved by further looking into the very
interesting results of Bozic \textit{et al}~\cite{Bozic_2004} in
two directions:

- by determining the particle "quantum trajectories" which take
into account their size and the slits size. Thus one obtains the
particles density after the slits which can be very different from
the square of the wave function.

- by proposing some experiment making it possible to clearly
highlight the difference between the density obtained by
calculation of the wave function and the density obtained by the
quantum trajectories.

We will make use of the experimental data of the Zeilinger team
corresponding to the $C_{60}$ molecule~\cite{Zeilinger_1999}: the
molecule is spherical with a diameter of 1 nm and the slits have a
width of 50 nm (it is the same size ratio as that of a soccer ball
compared to the goal).

Indeed, as Bozic \textit{et al}~\cite{Bozic_2004} point out, the
quantum description must theoretically take into account the
interaction of an extended particle with the edge of the slits.
The experiment with $C_{60}$~\cite{Zeilinger_1999} shows that one
can neglect this effect if the ratio of the sizes is rather
large($\sim 50$ nm). This is what we will do from now on. The
error could indeed come only from the particles which run up
against the edges, which according to the preceding estimate would
amount to an error of the order of a percent only.

Section 2 describes a thought experiment corresponding to
asymmetrical slits, then section 3 calculates the modulus square
of the wave function after the slits both of this experiment and
of some diffraction experiment. Then section 4 shows how the
Broglie-Bohm trajectories make it possible to calculate the
particles density after the slits. From that we conclude that the
thought experiment suggested is a crucial experiment to test the
Broglie-Bohm interpretation for indistinguishable particles.

\section{The thought experiment}

We propose a thought experiment inspired by the real experiment of
Arndt \textit{et al}~\cite{Zeilinger_1999}. One considers a
molecular beam of fullerenes $C_{60}$, whose speed is $v_y = 200
m/s$ along the (0y) axis. Initial speeds in the other directions
are considered null. The molecular beam is 7 $\mu$m wide along the
(0x) axis.

At $d_1=1$ meter of the orifice of the molecular beam a plate is
placed admitting a slit A of 50 nm along the 0x axis and a grating
B of 100 small slits of 0.5 nm of period 1 nm along the 0x axis.
The distance between the centers of A and B is of 150 nm. The
molecules of fullerenes are then observed by using a scanning
laser-ionization detector placed at $d_2=1.25$ m after the slits.

\begin{figure}[H]
\begin{center}
\includegraphics[width=.8\linewidth]{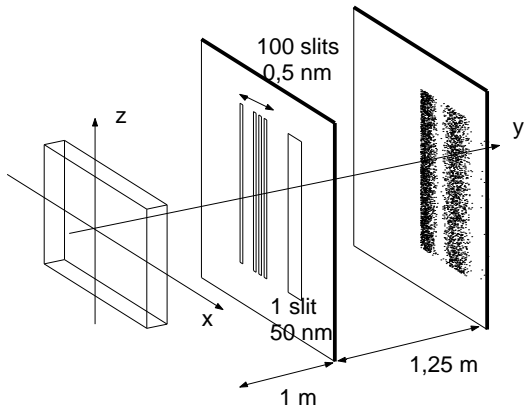}
\caption{\label{fig:schemexperience} Schematic diagram of the
thought experiment.}
\end{center}
\end{figure}

Slit A corresponds to the slits of the experiment of Arndt
\textit{et al}~\cite{Zeilinger_1999}. The size of the slits of
grating B was chosen to stop the molecules $C_{60}$. Currently,
such a grating is certainly impossible to realize to day, but it
will allow the thought experiment. It corresponds to the diagram
of an experiment which the developments in nanotechnology should
make possible. It is perhaps experimentally easier to construct a
hole of 50 nm diameter in the place of slit A and a set of 10000
small holes of 0.5 nm diameter in the place of grating B;  but
with slits the theoretical study is clearer and the simulations
are easier.

With the velocity $v_y = 200 m/s$, the mass of the fullerene
$C_{60}$, m=1.2$\times10^{-24}$ kg gives, a de Broglie wavelength
$\lambda_{db}=2.8$ pm, 350 times smaller than its diameter ($\sim$
1 nm).

We will compare this asymmetrical slits (slit A and grating B)
experiment with the diffraction experiment involving only slit A.

\section{Calculation of the wave function with Feynman path integral}

The calculation of the wave function is obtained by a numerical
calculation using the Feynman's integrals, as we
did~\cite{Gondran_2005} for the numerical simulation of the
experiment of the slits of Shimizu\textit{et al } with cold
atoms.\cite{Shimizu}

For the numerical simulation, we make the following assumptions.
The slits being very long along the 0z axis, there is no
diffraction according to this axis, but the particles are
subjected to gravity along (0z). Consequently, the variable z can
be treated classically as the varible y, satisfying the relations
$y=v_y t$ and $z=-\frac{1}{2}g t^2$. We thus consider only the
wave function in x, $\psi(x,t)$; we take as initial wave function
$\psi_0(x)=(2\pi \sigma_0^2)^{-\frac{1}{4}}\exp^{-\frac{x^2}{4
\sigma_0^2}}$ with $\sigma_0=2~\mu m$.

The wave function before the slits is then equal to
\begin{equation}\label{eq:eqavantfentes}
\psi(x,t)=(2\pi s(t)^2)^{-\frac{1}{4}}\exp^{-\frac{x^2}{4 \sigma_0
s(t)}}
\end{equation}
with $s(t)=\sigma_0 (1+\frac{i \hbar t}{2 m \sigma_0^2})$. After
the slits, at time $t\geq t_1=\frac{d_1}{v_y}=5~ms$, we use the
Feynman path integral method to calculate the time-dependent wave
function \cite{FeynmanQMI}~:
\begin{equation}\label{eq:eqapresfentes}
\psi(x,t)=\int_{F}K(x,t;x_f,t_1) \psi(x_f,t_1) dx_f
\end{equation}
where
\begin{equation}\label{eq:eqdeK}
K(x,t;x_f,t_1)=(\frac{m}{2 i \pi \hbar
(t-t_1)})^{\frac{1}{2}}\exp\frac{i
m}{\hbar}\frac{(x-x_f)^2}{2(t-t_1)}
\end{equation}
and where integration in (\ref{eq:eqapresfentes}) is carried out
on set F of the area of the various slits and where $\psi(x_f,
t_1)$ is given by (\ref{eq:eqavantfentes}).

On figure 2 is represented the modulus square of the wave function
on the detection screen of (a) the diffraction experiment with
slit A only, and (b) the interference experiment with asymmetrical
slits (slit A and grating B). Figure 2c is just a zoom on central
part of figure 2b. We note that 20 percent of the total density is
not represented on figure 2c, as one part moves laterally towards
the right of the screen and another part leftwards. For the
asymmetrical slits experiment, the modulus square of the wave
function is asymmetrical in the first centimeters after the slits,
then becomes fairly symmetrical when it comes to the detection
screen placed at 1.25 meter (figures 2b and 2c).

\begin{figure}[H]
\begin{center}
\includegraphics[width=0.4\linewidth]{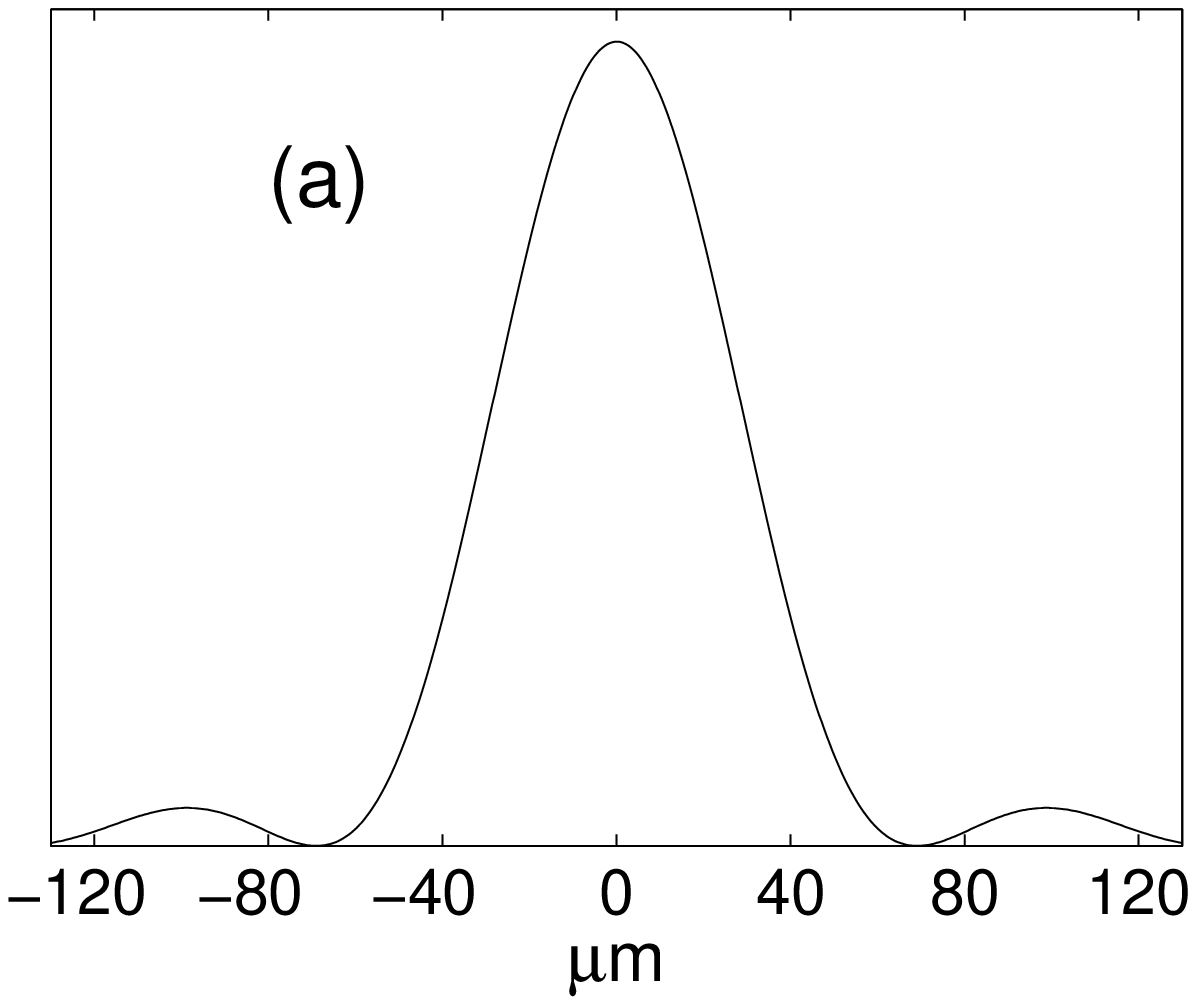}
\includegraphics[width=0.4\linewidth]{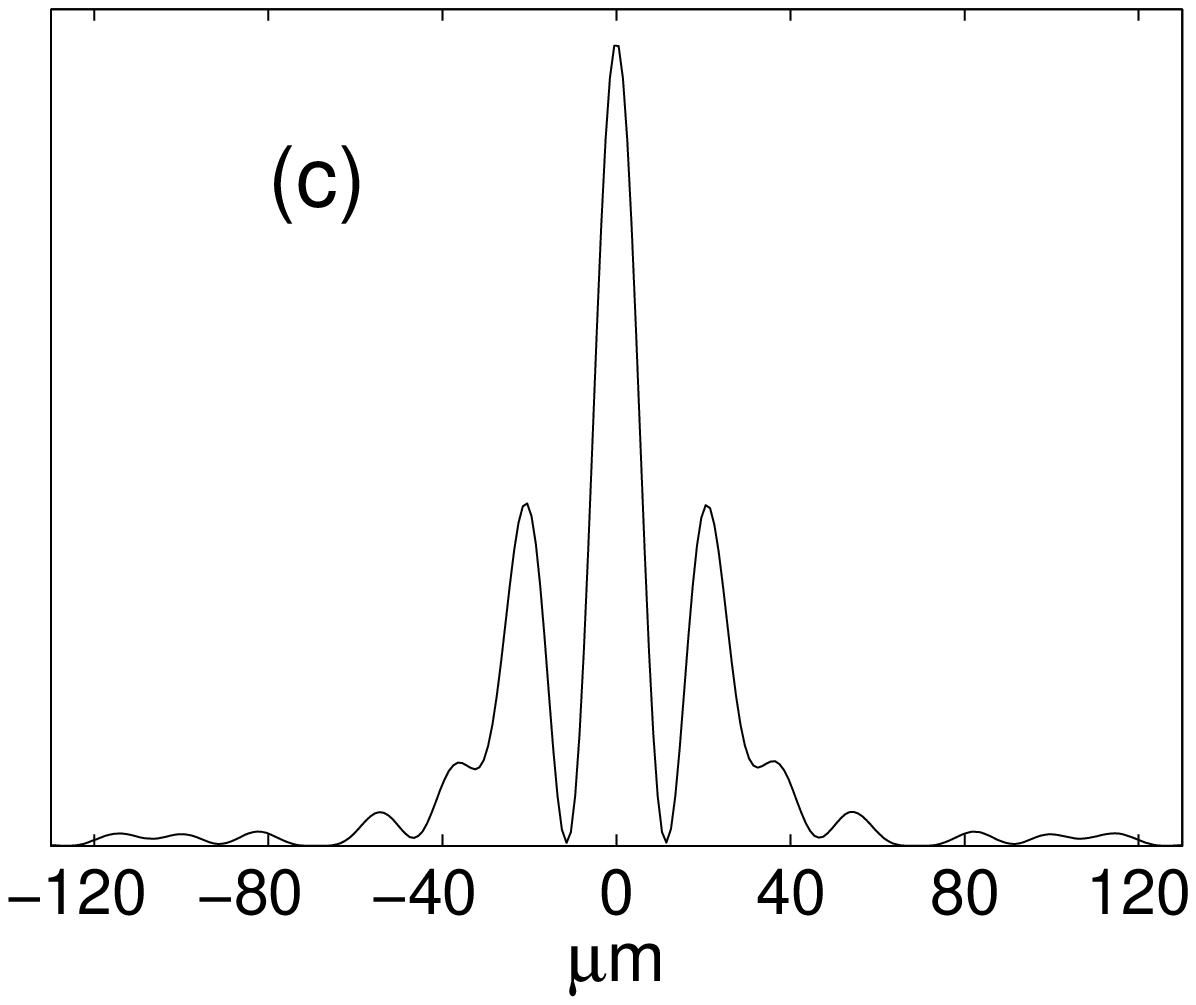}
\includegraphics[width=0.4\linewidth]{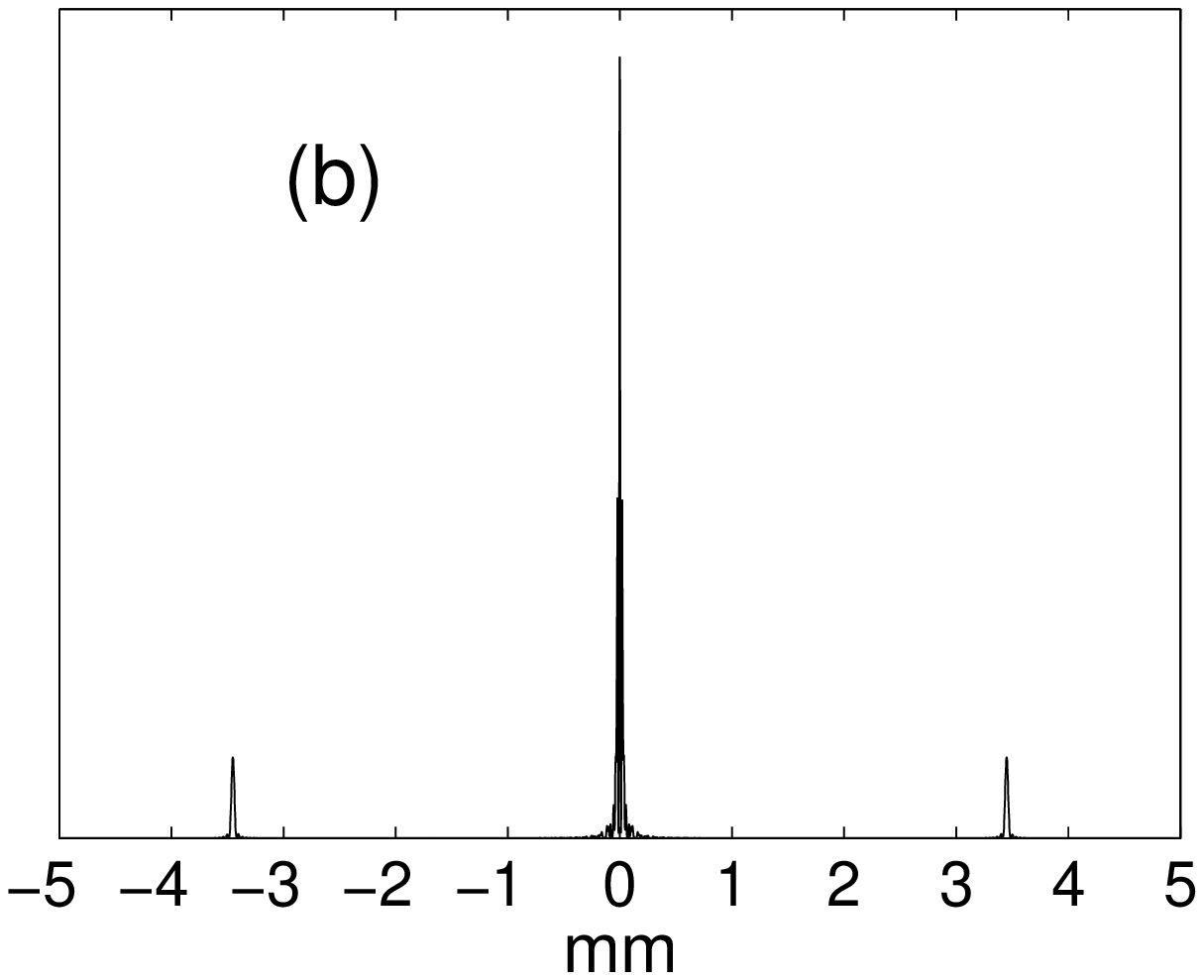}
\caption{\label{fig:comparaisonInterferenceDiffractions} Modulus
square of the wave function on the detection screen respectively:
(a) diffraction (slit A), (b) interference with asymmetrical slits
(slit A and grating B), (c) zoom of the asymetric slits central
part.}
\end{center}
\end{figure}

The standard interpretation of quantum mechanics postulates that
the density of the particles must be equal to the modulus square
of the wave function. The particles density on the detection
screen of the asymmetrical slits experiment must thus be given by
figures 2b and 2c (Standard Assumption 1): we obtain three peaks
on figure 2b (a central peak and two little peaks at -3,4 mm and
3,3 mm) and also three peaks on figure 2c (a central peak and two
smaller peaks at -20 $\mu m$ and +20 $\mu m$). If it is supposed
that the molecules cannot pass through grating B, the density
shown on figures 2b and 2c can then be in contradiction with the
experimental results. However within the framework of standard
interpretation, another solution is possible: one can make the
assumption that if the particles do not pass through grating B,
then the wave function does not pass through there either
(Standard Assumption 2). The experimental result must then be
given by figure 2a of the diffraction with the single slit A which
corresponds now to one unique peak.

\section{Calculation of the particles density with Broglie-Bohm trajectories}

In the Young slits experiments the interference fringes, and thus
the wave function, are never observed directly. The only direct
measurements are the individual impacts of the particles on the
detection screen. In the Broglie-Bohm
interpretation,~\cite{deBroglie,Bohm,BohmHolland} the particle is
represented not only by its wave function, but also by the
position of its center of mass. The center of mass of the particle
follows a trajectory, which is piloted by the wave function $\psi$
with a speed $\textbf{v}$ given by
\begin{equation}\label{eq:eqvitesseBohm}
\textbf{v}=\frac{\hbar}{2m \rho } Im{(\psi^\dag\boldsymbol\nabla
\psi)}.
\end{equation}
This interpretation statistically gives, in all the examples
available in the literature, the same experimental results as the
Copenhagen interpretation. Moreover, the Broglie-Bohm
interpretation naturally explains the individual impacts. These
impacts correspond, as in classic mechanics, with the position of
the particles at the time of their arrival on the detection
screen.

The experiment with asymmetrical slits corresponds to a case where
the Copenhagen and Broglie-Bohm interpretrations give different
results. It is then possible to test the Broglie-Bohm Assumption
in a very simple manner. One symply has to simulate the
Broglie-Bohm trajectories and to compare these simulation results
with the experiment.

In the standard interpretation, a molecule $C_{60}$ is represented
at the initial time only by its wave function $\psi_0(x)=(2\pi
\sigma_0^2)^{-\frac{1}{4}}\exp^{-\frac{x^2}{4 \sigma_0^2}}$. It
thus has an uncertain initial position since the wave function
only gives the probability density $\rho_0(x)=(2\pi
\sigma_0^2)^{-\frac{1}{2}}\exp^{-\frac{x^2}{2 \sigma_0^2}}$. Since
the particle is indistinguishable, at time $t$ one can only know
its probability density. Such indistinguishable particles can also
be found in classic mechanics, when one only knows speed and
initial probability density: to describe the evolution of such
classic particles and to make them distinguishable, it is
necessary to know their initial positions. In quantum mechanics,
de Broglie and Bohm make the same assumption for indistinguishable
particles.

As we did for cold atoms\cite{Gondran_2005}, we set up a Monte
Carlo simulation of the experiment by randomly drawing the initial
positions for molecules $C_{60}$ in the initial wave function.

\begin{figure}[H]
\begin{center}
\includegraphics[width=0.45\linewidth]{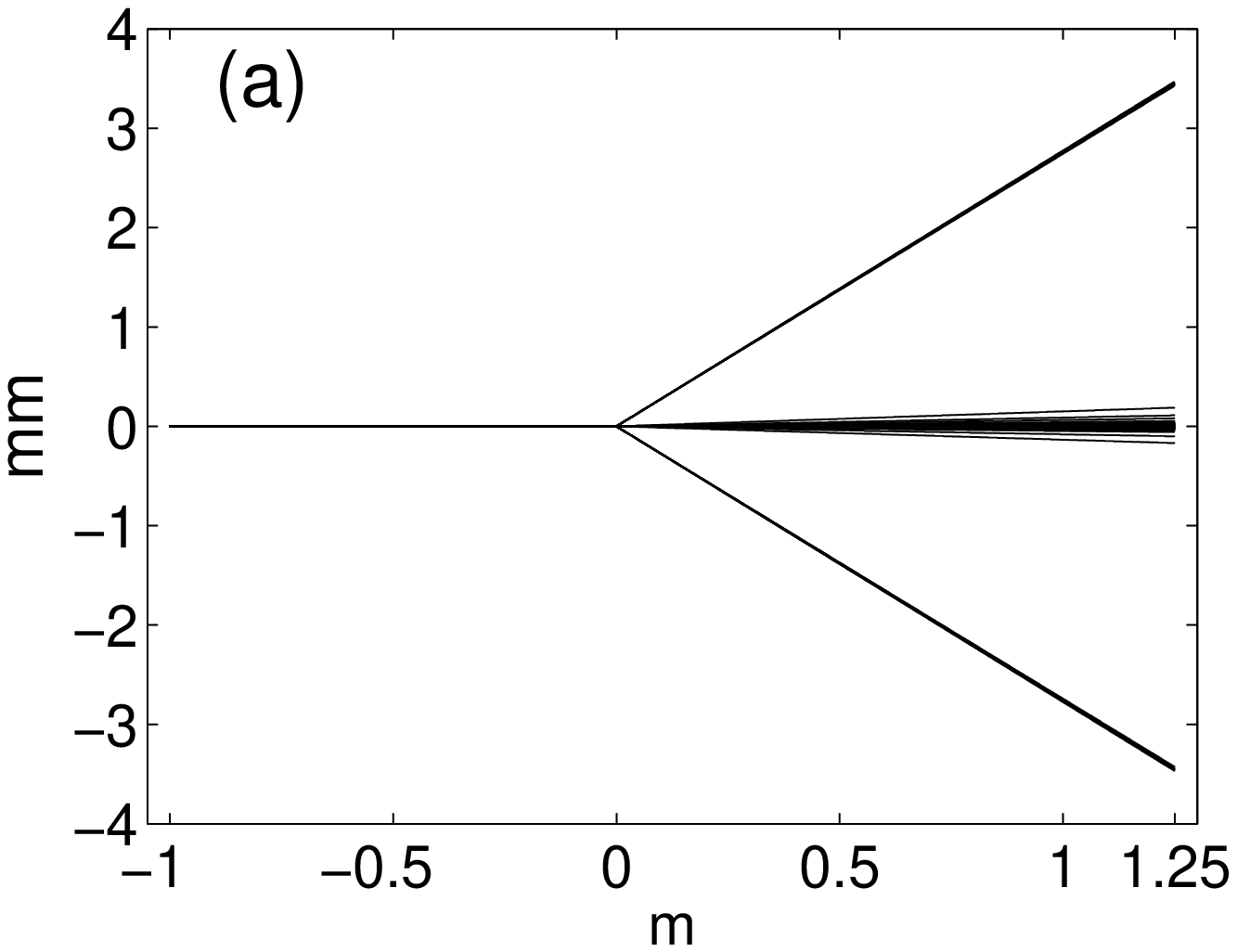}
\includegraphics[width=0.45\linewidth]{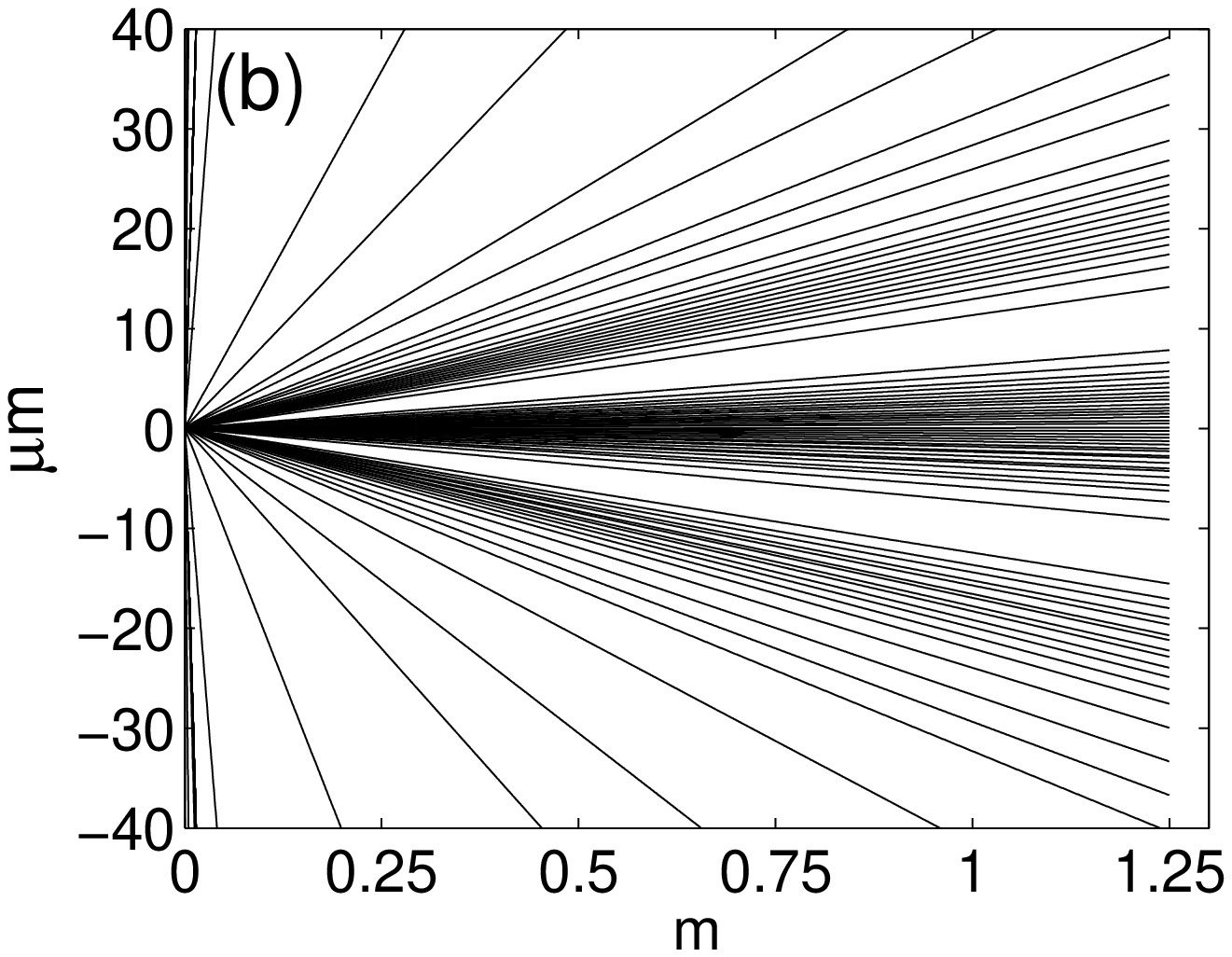}
\includegraphics[width=0.45\linewidth]{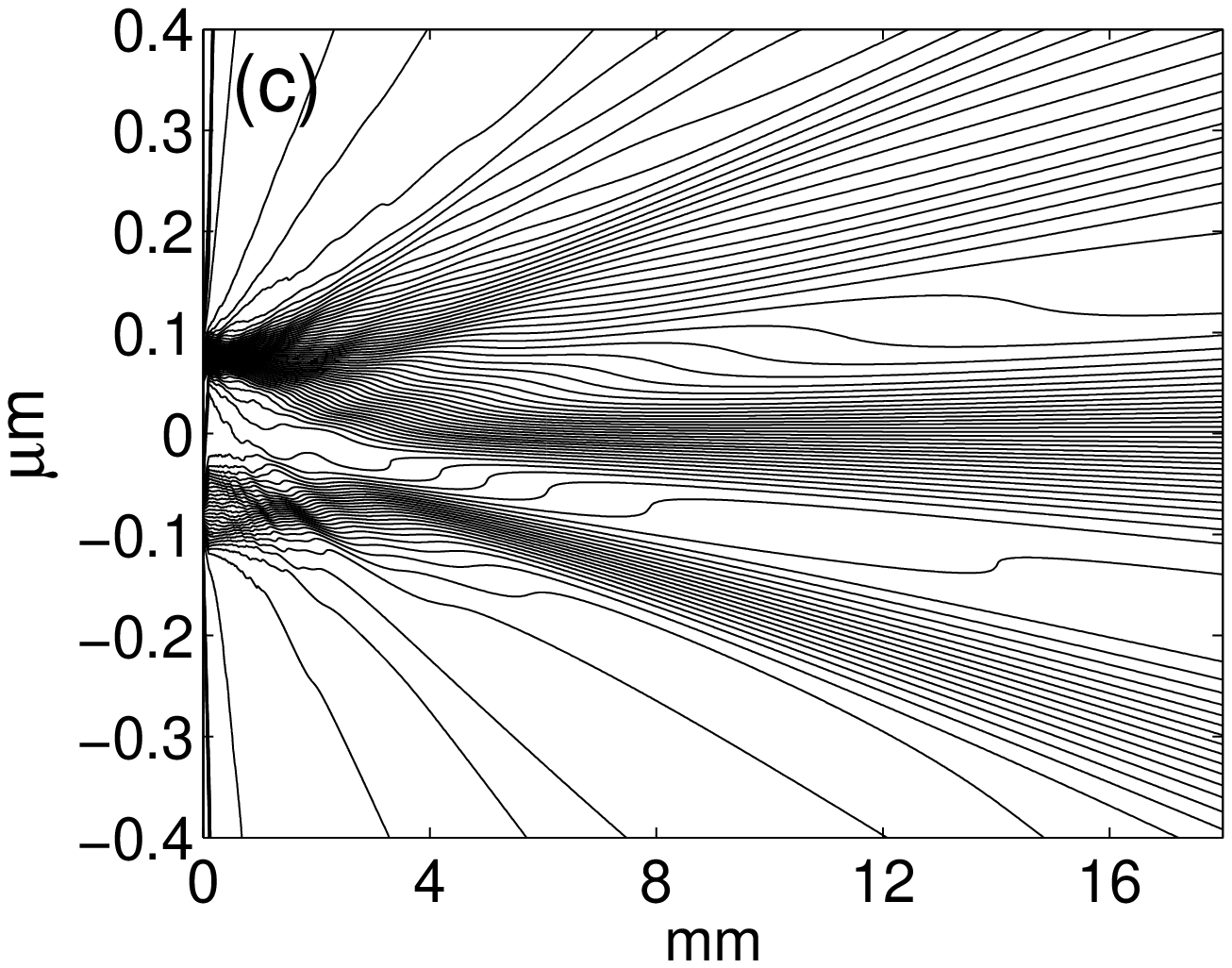}
\includegraphics[width=0.45\linewidth]{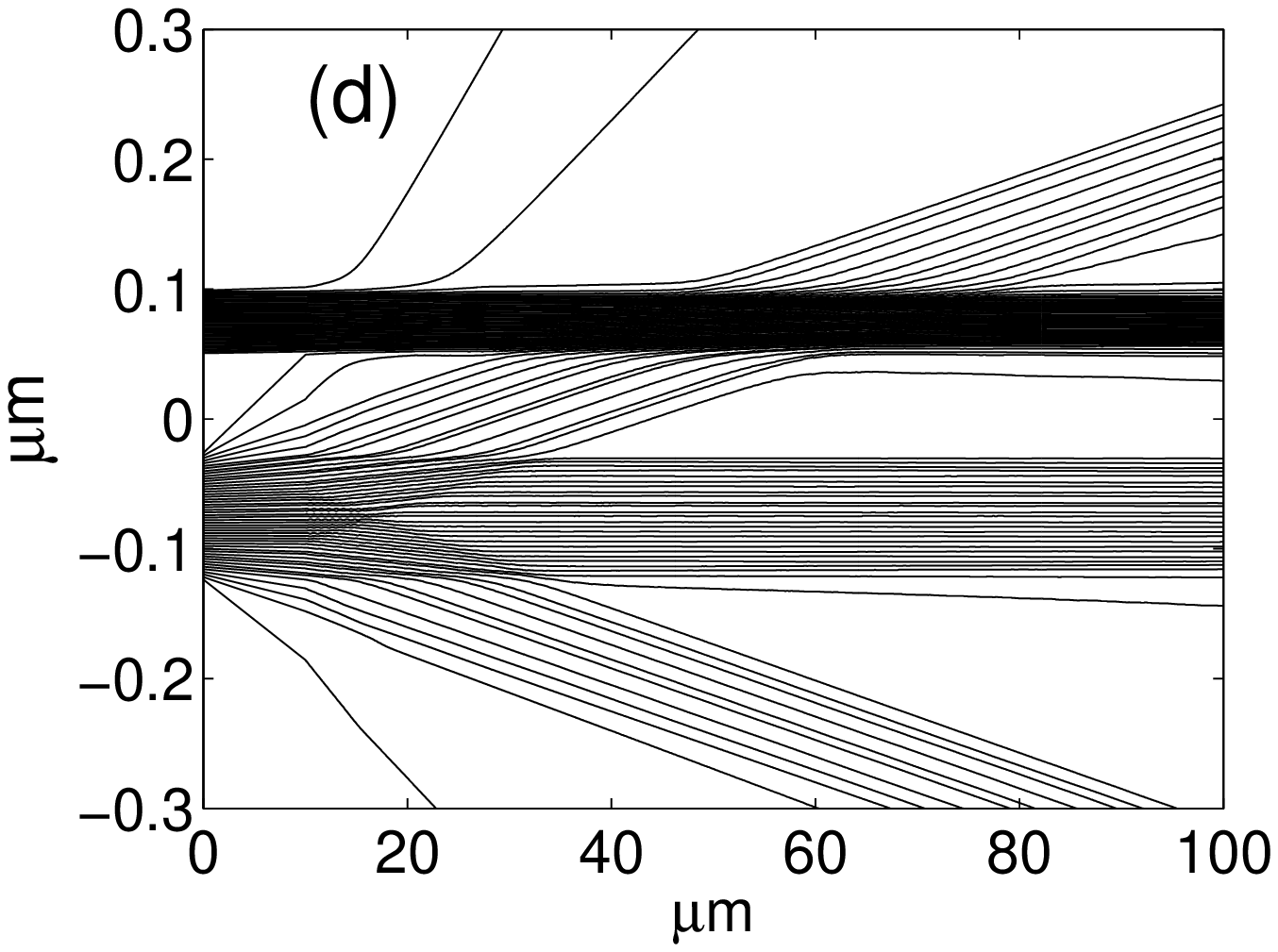}
\caption{\label{fig:} 100 Broglie-Bohm trajectories with randomly
drawn initial positions: (a) global view, (b) central
trajectories, (c) zoom on the first millimeters after the slits,
(d) zoom on the hundred first micrometers after the slits.}
\end{center}
\end{figure}

In the first place, one does not take into account the $C_{60}$
size. Figure 3 represents the quantum trajectories of 100
molecules $C_{60}$ which cross either slit A or grating B (one did
not represent the molecules stopped by the plate). The density of
these molecules on the detection screen corresponds to the modulus
square of the wave function represented on figures 2b and 2c. On
figure 3d we retrieve the loss of 20 percent of the density in the
trajectories, which accounts for particles which move laterally.
At a second stage one takes into account the fact that all
molecules $C_{60}$ are stopped by grating B. Figure 4 shows the
quantum trajectories of molecules $C_{60}$ which pass slit A only.
\begin{figure}[H]
\begin{center}
\includegraphics[width=0.45\linewidth]{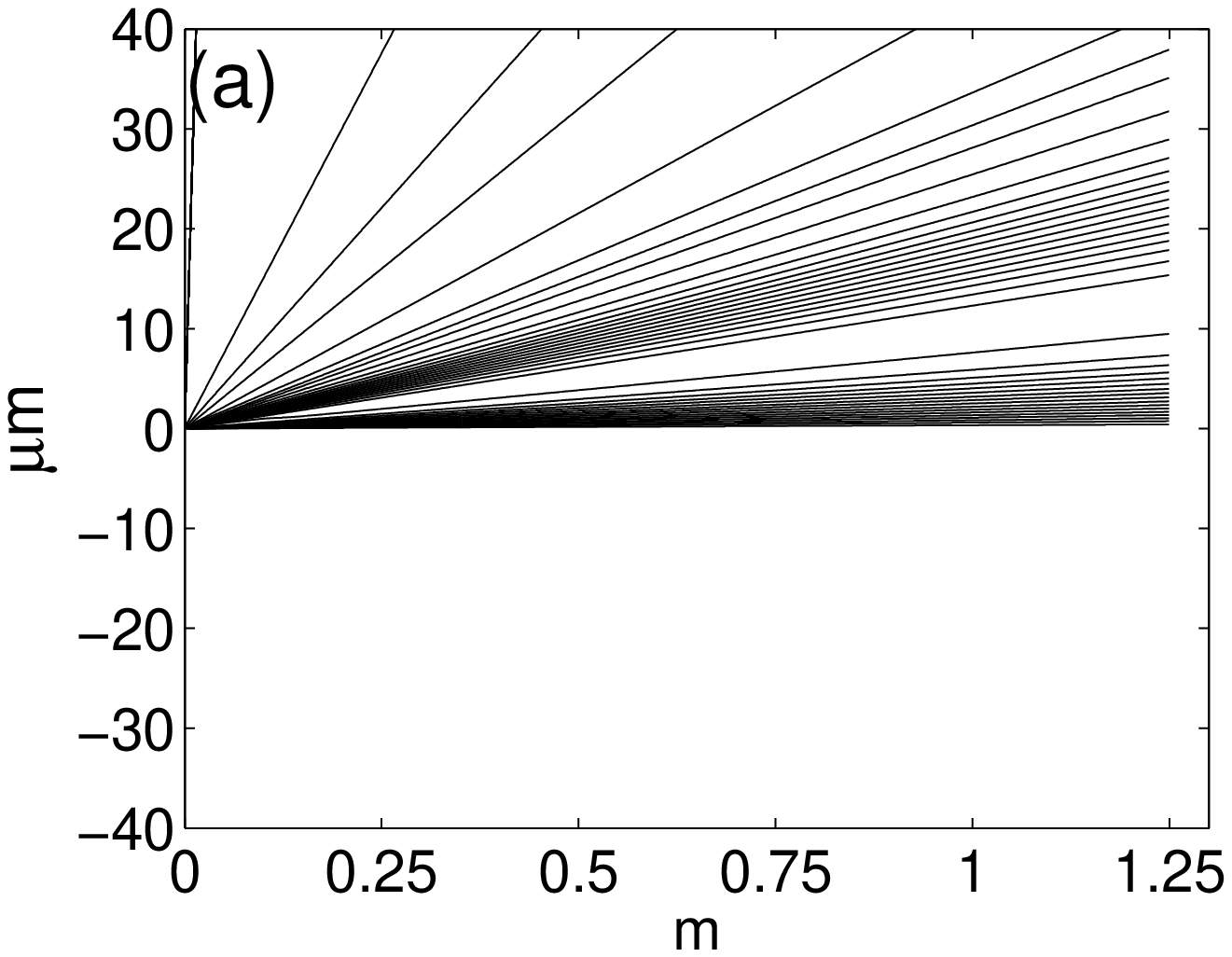}
\includegraphics[width=0.5\linewidth]{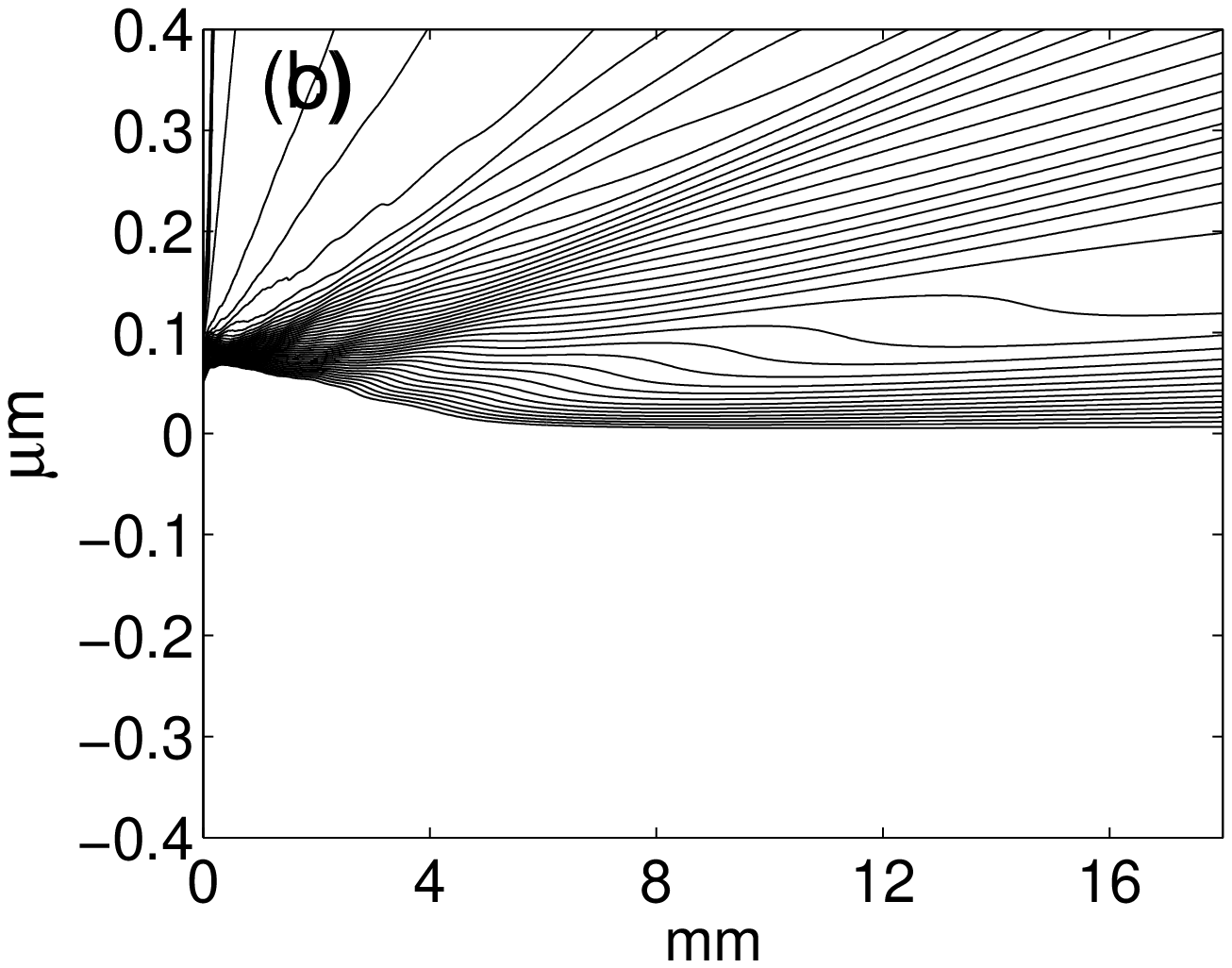}
\includegraphics[width=0.45\linewidth]{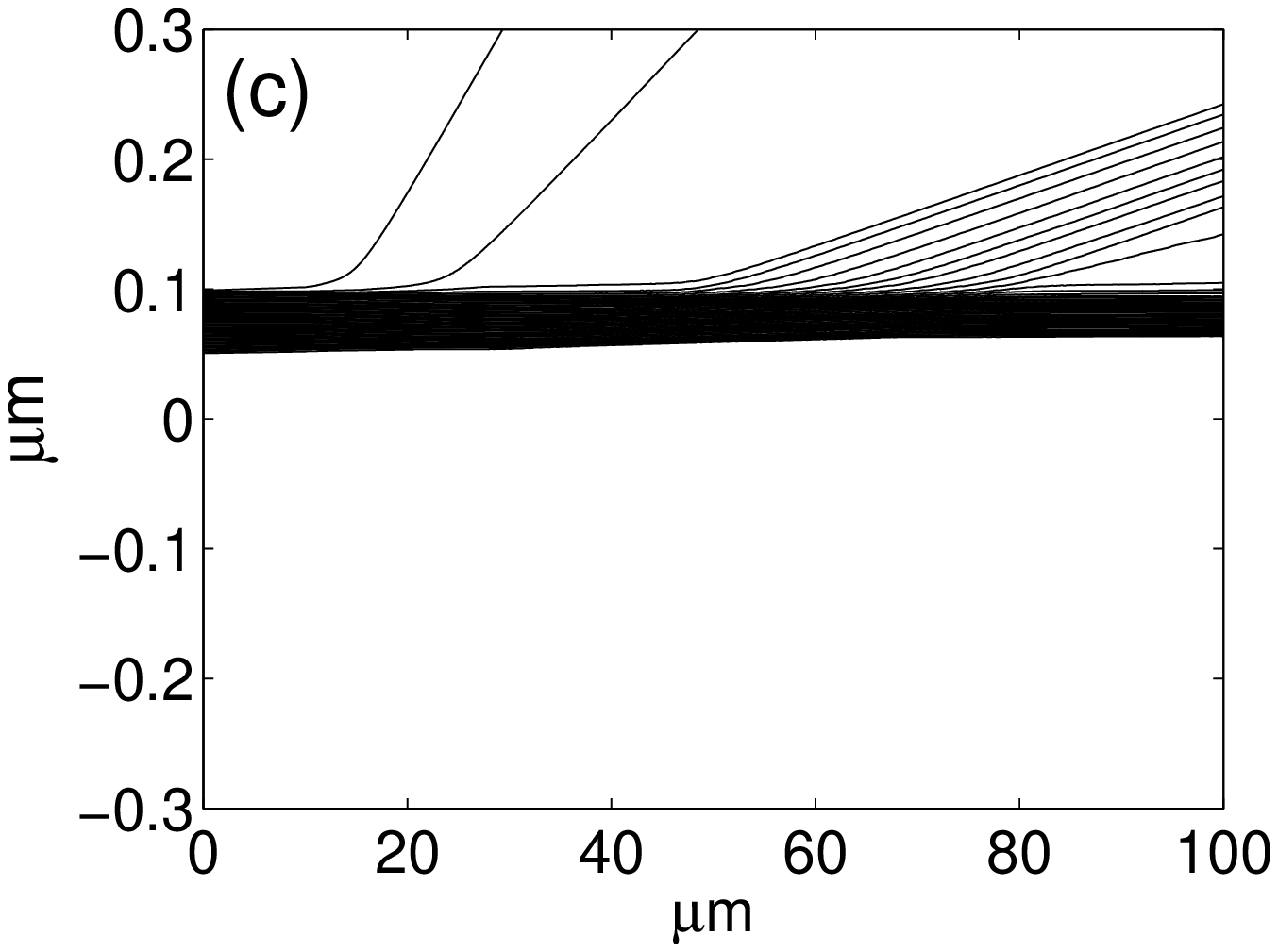}
\caption{\label{fig:densitedesmolecules} Broglie-Bohm trajectories
through slit A only: (a) global view of trajectories , (b) zoom on
the first millimeters, (c) zoom on the hundred first micrometers.}
\end{center}
\end{figure}
Figure 5 shows the density on the detection screen of molecules
$C_{60}$ which pass slit A only. The density is to be found
experimentally if the Broglie-Bohm assumption is valid.
\begin{figure}[H]
\begin{center}
\includegraphics[width=0.4\linewidth]{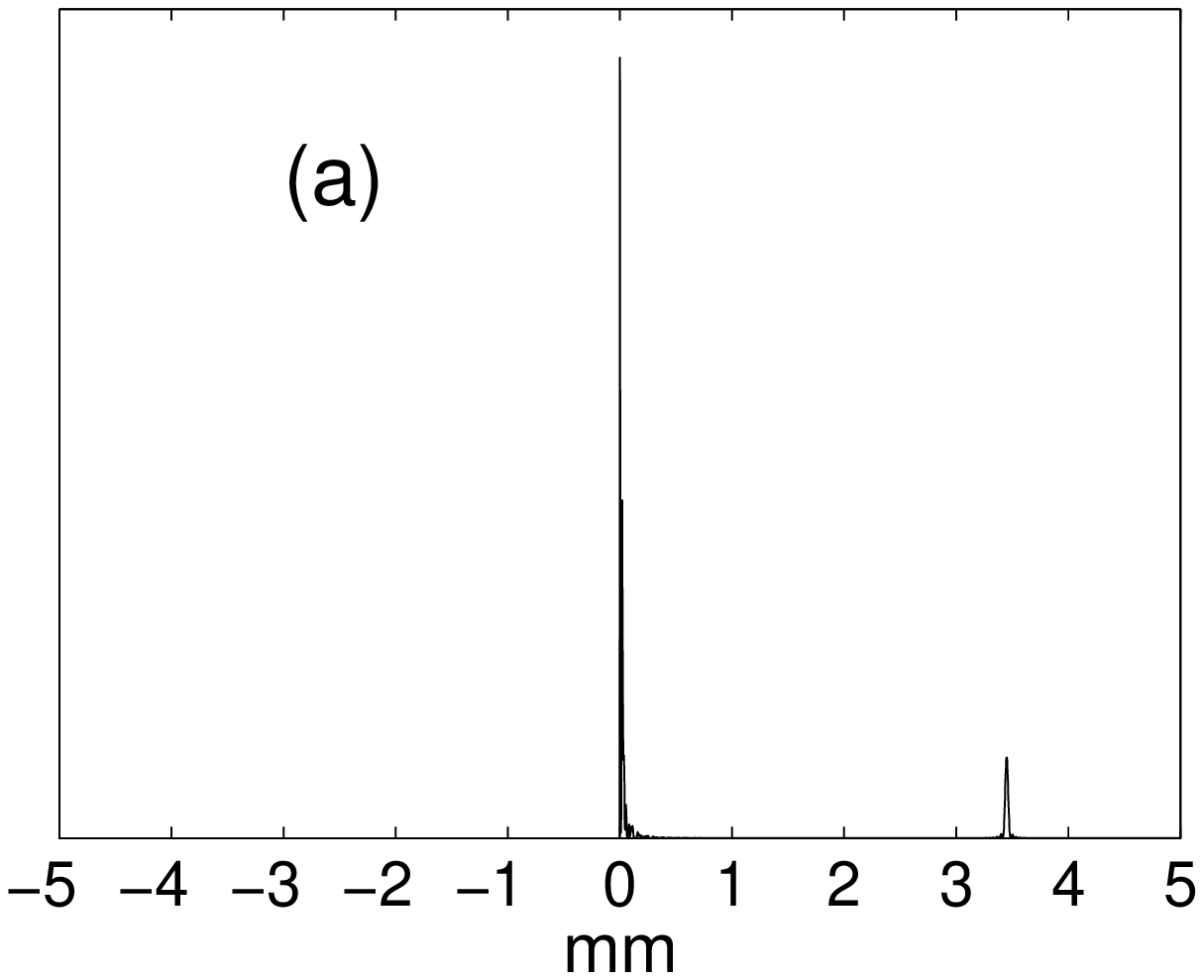}
\includegraphics[width=0.4\linewidth]{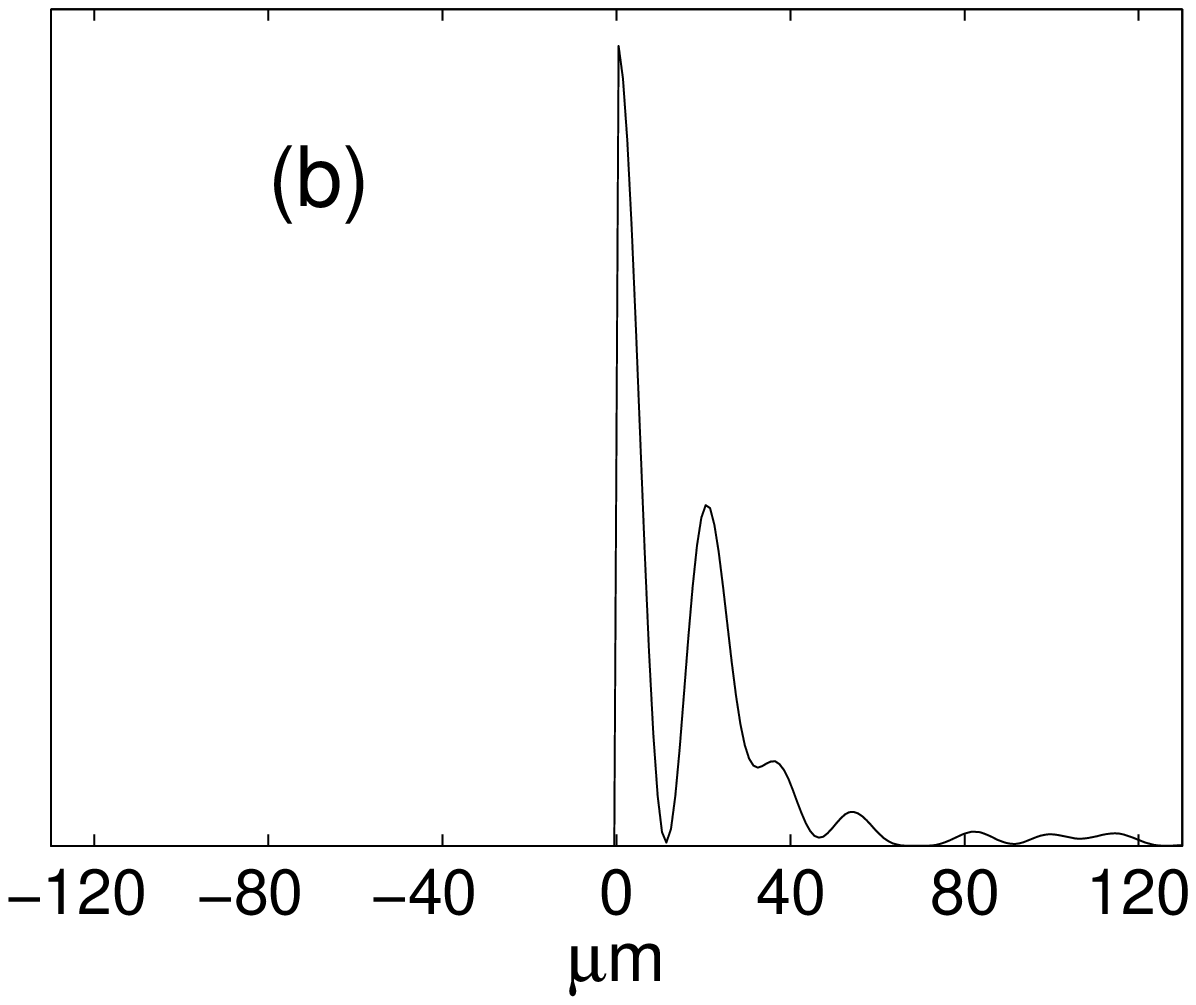}
\caption{\label{fig:densitedesmolecules2} Density of molecules
$C_{60}$ having crossed slit A: (a) global density, (b) central
density.}
\end{center}
\end{figure}

The clear difference between the density of Standard Assumption 1
(figures 2b and 2c with three peaks), Standard Assumption 2
(figure 2a with one peak) and the Broglie-Bohm Assumption
(asymmetrical figures 5a and 5b with two peaks only) shows that it
is possible to define a robust test in spite of certain
experimental difficulties:

- in preceding calculations we supposed that initial speeds were
well-known. It is not the case in the experiments of Zeilinger
where for example speed $v_y$ is not well-known. We have shown
\cite{Gondran_2005} how to take these uncertainties into account.
They will smooth the densities, but will preserve the number of
peaks.

- to ensure the validity of the calculation of wave function (2),
it is necessary to prevent molecules $C_{60}$ from being blocked
in the slits of grating B and stopped there. For that purpose, one
can slightly incline the plate so that the particles rebound while
falling downwards. One can also send them one by one in order to
leave to those which are stopped enough time to fall before the
arrival of the following particles. One can also make slits even
narrower.


\begin{figure}[H]
\begin{center}
\end{center}
\end{figure}

\bigskip

\section{Conclusion}

Taking into account the size of the particles in the interference
phenomena, makes it possible to renew the study of the
wave-particle duality and to propose experiments to test the
Broglie-Bohm trajectories for indistinguishable particles. Should
the test be positive, the wave function for indistinguishable
particles would have to be considered as a field. For the
distinguishable particles, this interpretation does not
apply~\cite{Gondran_2006}.

While waiting for the results of this test which raises
experimental difficulties, one can already realize a much simpler
experiment which makes it possible to clarify the standard
postulate of the probabilistic interpretation of the modulus
square of the wave function: it is the test of Standard Assumption
2 against Standard Assumption 1. This can be achieved
experimentally by simply reducing the size of slits and holes to
less than the diameter of molecule $C_{60}$. That should suffice
to observe that the density on the detection screen is null.

Professor Bozic has suggested that our thougt experiment might be
carried out with Rydberg atoms instead of $C_{60}$ molecules.
Indeed Fabre et al~\cite{Fabre_1983} have shown that slits of 1
$\mu m$ are enough to stop Rydberg atoms as soon as $n=50$. The
experiment would then be performable right now.

\section*{Acknowledgements}
We are very grateful to Professor Bozic for having given us access
to his work on asymmetrical slits and for the discussions we had
on this subject.

\end{document}